\documentstyle{mn}
\input{epsf}
\newcommand{\EQ}{\begin{equation}}
\newcommand{\EN}{\end{equation}}
\newcommand{\EQA}{\begin{eqnarray}}
\newcommand{\ENA}{\end{eqnarray}}

\newcommand{\Eq}[1]{Eq.~(\ref{#1})}
\newcommand{\Eqs}[2]{Eqs~(\ref{#1}) and~(\ref{#2})}
\newcommand{\Eqss}[2]{Eqs~(\ref{#1})--(\ref{#2})}

\newcommand{\Sec}[1]{Sect.~\ref{#1}}
\newcommand{\Fig}[1]{Fig.~\ref{#1}}

\newcommand{\Figs}[2]{Figs~\ref{#1} and \ref{#2}}

\newcommand{\bra}[1]{\langle #1\rangle}

\newcommand{\meanAA}{\overline{\mbox{\boldmath $A$}}}
\newcommand{\meanBB}{\overline{\mbox{\boldmath $B$}}}
\newcommand{\meanJJ}{\overline{\mbox{\boldmath $J$}}}

\newcommand{\eee}{\hat{\mbox{\boldmath $e$}} {}}

\newcommand{\yyy}{\hat{\mbox{\boldmath $y$}} {}}

\newcommand{\xx}{\mbox{\boldmath $x$} {}}

\newcommand{\uu}{\mbox{\boldmath $u$} {}}

\newcommand{\bb}{\mbox{\boldmath $b$} {}}

\newcommand{\BB}{\mbox{\boldmath $B$} {}}

\newcommand{\AAA}{\mbox{\boldmath $A$} {}}

\newcommand{\jj}{\mbox{\boldmath $j$} {}}
\newcommand{\JJ}{\mbox{\boldmath $J$} {}}

\newcommand{\ff}{\mbox{\boldmath $f$} {}}

\newcommand{\kk}{\mbox{\boldmath $k$} {}}

\newcommand{\nab}{\mbox{\boldmath $\nabla$} {}}

\newcommand{\oo}{\mbox{\boldmath $\omega$} {}}

\newcommand{\DD}{{\rm D} \, {}}
\newcommand{\dd}{{\rm d} {}}

\newcommand{\ea}{{\rm et al. }}

\def\onethird{{\textstyle{1\over3}}}

\newcommand{\yr}{\,{\rm yr}}

\newcommand{\yjgr}[3]{ #1, {JGR }{#2}, #3}
\newcommand{\yapj}[3]{ #1, {ApJ }{#2}, #3}
\newcommand{\yapjl}[3]{ #1, {ApJ (Letters) }{#2}, #3}

\newcommand{\yana}[3]{ #1, {A\&A }{#2}, #3}

\newcommand{\ypf}[3]{ #1, {Phys. Fluids }{#2}, #3}

\newcommand{\ynat}[3]{ #1, {Nat }{#2}, #3}

\newcommand{\ysph}[3]{ #1, {Solar Phys. } {#2}, #3}
\newcommand{\ypr}[3]{ #1, {Phys. Rev. } {#2}, #3}

\newcommand{\ybook}[3]{ #1, {#2} (#3)}

\newcommand{\sana}[1]{ #1, {A\&A} (submitted)}

\newcommand{\sgafd}[1]{ #1, {GAFD} (submitted)}

\begin{document}
\title{The helicity constraint in turbulent dynamos with shear}
\author[A.\ Brandenburg, A.\ Bigazzi and K.\ Subramanian]{Axel Brandenburg$^{1,2}$,
Alberto Bigazzi$^{3}$, and Kandaswamy Subramanian$^4$\\
$^1$ NORDITA, Blegdamsvej 17, DK-2100 Copenhagen \O, Denmark\\
$^2$ Department of Mathematics, University of Newcastle upon Tyne, NE1 7RU, UK\\
$^3$ Department of Mathematics, Politecnico di Milano,
Piazza Leonardo da Vinci 32, I-20133 Milano, Italy\\
$^4$ National Centre for Radio Astrophysics - TIFR, Poona University Campus,
Ganeshkhind, Pune 411 007, India}

\maketitle

\begin{abstract}
The evolution of magnetic fields is studied using simulations of
forced helical turbulence with strong imposed shear. After some initial
exponential growth, the magnetic field develops a large scale travelling
wave pattern. The resulting field structure possesses magnetic helicity,
which is conserved in a periodic box by the ideal MHD equations and
can hence only change on a resistive time scale. This constrains strongly
the growth time of the large scale magnetic field, but less strongly the
length of the cycle period.
Comparing with the case without shear, the time scale for
large scale field amplification is shortened by a factor $Q$, which
depends on the relative importance of shear and helical turbulence,
and which controls also the ratio of toroidal to poloidal field.
The results of the simulations can be reproduced qualitatively and
quantitatively with a mean-field $\alpha\Omega$ dynamo model with
alpha-effect and the turbulent magnetic diffusivity coefficients that
are less strongly quenched than in the corresponding $\alpha^2$-dynamo.

\end{abstract}

\section{Introduction}

In astrophysical bodies such as stars and galaxies there is a strong
magnetic field of large scale. Such fields have usually significant
magnetic helicity (e.g., Pevtsov \ea 1995, Berger \& Ruzmaikin 2000).
This is non-trivial, because magnetic helicity is a
conserved quantity and can only change if there is a flux of helicity
through the boundaries, or through resistive effects which are however
very slow. Although this has been known for some time, it is only recently that
this has been identified as the fundamental reason for `catastrophic'
quenching of the $\alpha$-effect in mean-field dynamo theory (Blackman
\& Field 2000, Kleeorin \ea 2000). Simulations of non-mirror symmetric
turbulence, which is prototypical of flows producing $\alpha^2$-dynamos,
have shown that a large scale helical magnetic field can only grow to
its final (super-) equipartition field strength on a {\it resistive}
time scale (Brandenburg 2001, hereafter referred to as B2001).

One may be tempted to sweep the problem of helicity conservation under
the carpet, because it has mainly been discussed in connection with
rather idealised models. We believe however that the problem is serious
and quite general. In fact, it also applies to convection-driven dynamos
and even to the case where the dynamo-generating flow is the result of
magnetic instabilities, as was found to be the case in simulations of
accretion discs with a dynamo-generated large scale field (Brandenburg
\ea 1995). This may be particularly surprising in view of the rather
plausible expectation that the $\alpha$-effect and turbulent diffusivity
should be `anti-quenched' and increase with increasing field strength
(Hasler \ea 1995, Brandenburg \ea 1998). If such a mechanism is to be successful, it must
still obey helicity conservation and can hence only produce a field with
vanishing net magnetic helicity.

There is strong observational evidence that the solar magnetic field is
indeed helical (Seehafer 1990, Pevtsov \ea 1995). These observations
suggest negative current helicity of the small scale fields in the
northern hemisphere. Using a relation by Keinigs (1983), this implies
a positive $\alpha$-effect (Seehafer 1996), which is
consistent with B2001. In order to produce finite net-helicity
one must get rid of fields with opposite sign of magnetic helicity, either
through dissipation (which is slow) or through selective losses through open
boundaries. So far there is no evidence from simulations however that
such losses involve fields of significant strength and opposite sign
of magnetic helicity relative to those that remain in the dynamo-active
domain (Brandenburg \& Dobler 2001).

The dynamo simulations that allowed addressing the question of the helicity
constraint were all of $\alpha^2$-type, so there was no additional field
amplification by shear. Thus, an outstanding question is therefore whether or
not the helicity constraint also plays a role in the presence of shear
through which strong toroidal magnetic fields can be generated without
affecting the magnetic helicity.

There are a number of working dynamos which have both open boundaries
and shear (e.g., Glatzmaier \& Roberts 1995, Brandenburg \ea 1995),
but those models are rather complex and use subgrid scale modelling,
so one cannot straightforwardly define an effective magnetic Reynolds
number. This makes a reliable assessment of the effects of helicity
conservation difficult. Nevertheless, it clearly remains one of the next
important tasks to reconsider these or similar simulations in the light of
helicity conservation. In order to determine the relative importance of
the various possibilities for relaxing the helicity constraint (shear,
open boundaries, etc.) it is useful to consider each possibility
in isolation. As a straightforward extension of the model of B2001
we consider here the inclusion of large scale sinusoidal shear, which
allows us to retain the assumption of periodic boundary conditions.

We have mentioned already that shear could be important for relaxing the
helicity constraint because the toroidal field generated by stretching
does not need to be helical and would hence not be subject to the
helicity constraint. On the other hand, shear alone is insufficient for
dynamo action: one needs an additional effect that regenerates poloidal
(cross-stream) field from toroidal field (e.g.\ Moffatt 1978, Krause \&
R\"adler 1980). The main point of the present paper is to show that, even
though much of the magnetic field amplification is due to shear, which
causes the field to be only weakly helical, the magnetic field is still
subject to a (modified) helicity constraint. More specifically, we shall
show that it is no longer the large scale
field as such which grows resistively, but rather the {\it geometrical
mean} of the magnitudes of the poloidal and toroidal mean fields. The
reason is simple: large scale helicity measures essentially the linkage
of poloidal and toroidal fields and must therefore be proportional to the
product of the two. The constraint
that helicity can change only on a resistive time scale can then be 
alleviated somewhat. This is because, now, for the same magnetic helicity,
stronger toroidal fields are possible at the expense of weaker poloidal
fields. Or conversely, equipartition strength large
scale fields can be attained in times shorter by the ratio of toroidal
to poloidal field strength.

\section{The model}
\label{Smodel}

As in B2001 we adopt the MHD equations for an isothermal compressible
gas, driven by a given body force $\ff$, which represents both shear
and small scale driving;
\EQ
{\DD\ln\rho\over\DD t}=-\nab\cdot\uu,
\EN
\EQ
{\DD\uu\over\DD t}=-c_{\rm s}^2\nab\ln\rho+{\JJ\times\BB\over\rho}
+{\mu\over\rho}(\nabla^2\uu+\onethird\nab\nab\cdot\uu)+\ff,
\label{dudt}
\EN
\EQ
{\partial\AAA\over\partial t}=\uu\times\BB
-\eta\mu_0\JJ,
\label{dAdt}
\EN
where ${\rm D}/{\rm D}t=\partial/\partial t+\uu\cdot\nab$ is the
advective derivative, $\uu$ is the velocity, $\rho$ is the density,
$\BB=\nab\times\AAA$ is the magnetic field, $\AAA$ is its vector
potential, $\JJ=\nab\times\BB/\mu_0$ is the current density, $\eta$ is
the magnetic diffusivity, and $\mu$ the dynamical viscosity. We adopt
a forcing function $\ff$ of the form
\EQ
\ff=\ff_{\rm turb}+\ff_{\rm shear},
\EN
where
\EQ
\ff_{\rm shear}=C_{\rm shear}{\mu\over\rho}\,\yyy\sin x
\EN
balances the viscous stress once a sinusoidal shear flow has been
established, and
\EQ
\ff_{\rm turb}=\mbox{Re}\{N\ff_{\kk(t)}\exp[i\kk(t)\cdot\xx+i\phi(t)]\},
\EN
is the small scale helical forcing with
\EQ
\ff_{\kk}={\kk\times(\kk\times\eee)-i|\kk|(\kk\times\eee)
\over2\kk^2\sqrt{1-(\kk\cdot\eee)^2/\kk^2}}.
\EN
Here $\eee$ is an arbitrary unit vector needed in order to generate
a vector $\kk\times\eee$ that is perpendicular to $\kk$, $\phi(t)$
is a random phase, and $N=f_0 c_{\rm s}(kc_{\rm s}/\delta t)^{1/2}$,
where $f_0$ is a nondimensional factor, $k=|\kk|$, and $\delta t$ is the
length of the time step. As in B2001 we focus on the case where $|\kk|$
is around $k_{\rm f}\equiv5$, and select at each time step randomly one
of the 350 possible vectors in $4.5<|\kk|<5.5$.

We use nondimensional units where $c_{\rm s}=k_1=\rho_0=\mu_0=1$. Here,
$c_{\rm s}$ is the sound speed, $k_1$ is the smallest wavenumber in
the box (so its size is $2\pi$), $\rho_0$ is the mean density (which is
conserved), and $\mu_0$ is the vacuum permeability.

We are interested in the case where shear is strong compared with the
turbulence, but still subsonic. In B2001 we used $f_0=0.1$ and found
that the resulting Mach number of the turbulence was between 0.1 and
0.3, which is already so close to unity that there would be no room to
accommodate sufficiently large shear which is still subsonic. Thus, we
now choose $f_0$ to be ten times smaller, and we take $f_0=0.01$. During
the saturated phase of the dynamo the resulting rms velocities in the
meridional ($xz$) plane are now around 0.015. For the shear parameter
we choose $C_{\rm shear}=1$, which leads to toroidal rms velocities
of around 0.6, which is about 40 times stronger than the velocities in
the meridional plane. The rms velocity from wavenumbers $k\ge2$ is
0.035, and this is also the value that we shall use for our estimates
of the magnetic Reynolds number and the equipartition field strength.

We choose a magnetic Prandtl number of 10, i.e.\
$\mu/(\rho_0\eta)=10$, and use $\eta=5\times10^{-4}$, so the magnetic
Reynolds numbers based on the box size ($=2\pi$) is about 400.
The magnetic Reynolds number based on the forcing scale is about 80.
The kinetic Reynolds number
based on the forcing scale is only 8, so one cannot expect
a proper inertial range. The turnover time based on the forcing scale
is $\tau=40$. In the following we denote
by poloidal and toroidal components those in the $xz$-plane and the
$y$-direction, respectively.

As usual for these type of simulations with helical forcing, there
is strong dynamo action at small scales amplifying an initially weak
random seed magnetic field exponentially (on a dynamical time scale) to
equipartition with kinetic energy. The poloidal field, which is strongly
dominated by small scales, saturates early on (at $t\approx1000$)
at a level of about 0.010--0.015. The toroidal field saturates later
(at $t\approx2000$) at a level of about 0.2--0.3, and is then already
dominated by large scales.

We begin by discussing the resulting field structure at late times,
turn then to the question of resistively limited growth of the {\it
large scale} field, and finally make comparisons with $\alpha\Omega$
dynamo theory.

\section{Field structure}

\begin{figure}\epsfbox{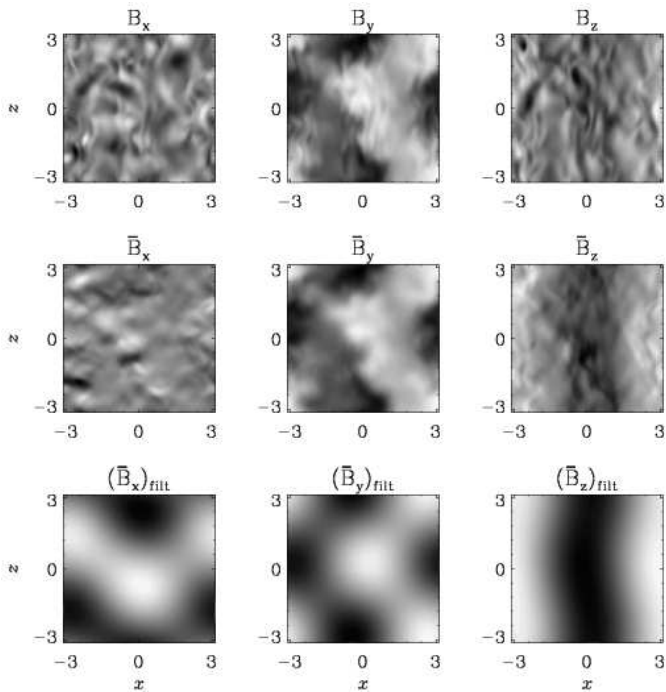}\caption[]{
Images of the three components of $\BB$ in an arbitrarily chosen $xz$ plane
(first row), compared with the $y$-averaged fields (second row) and the
fourier-filtered $y$-averaged fields with $|\kk|\leq2$, indicated by the
subscript `filt' (third row). $120^3$ meshpoints, $t=6000$.
}\label{Fpbmer}\end{figure}

In \Fig{Fpbmer} we show images of the three field components in the
meridional plane. Note that the toroidal field shows much smoother and
larger scale structures than the meridional field components. Moreover,
the toroidal field shows almost no variation along the $y$-directions:
the toroidal average, ${\overline B}_y$, (second row), is very similar
to an individual meridional cross-section of $B_y$, first row. However,
in contrast to the case without shear, where the mean fields showed
systematic variations only in one of the three coordinate directions
(B2001), here the toroidal field varies with both $x$ and $z$, consisting
of a superposition of modes with $k_x=1$ and $k_z=1$.

\epsfxsize=8.9cm

\begin{figure}\epsfbox{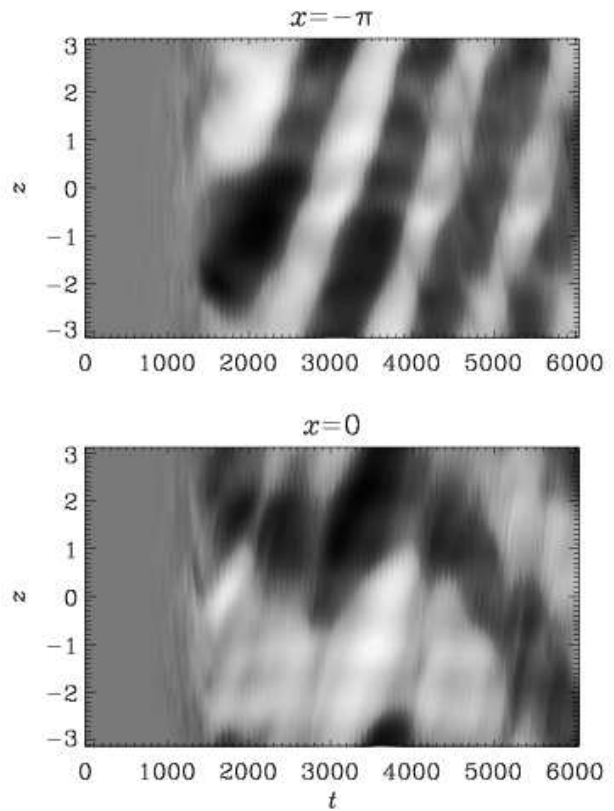}\caption[]{
Space-time diagram of the mean toroidal field at $x=-\pi$ (negative
local shear) and $x=0$ (positive local shear). Dark (light) shadings
refer to negative (positive) values. Note the presence of dynamo
waves travelling in the positive (negative) $z$-direction for negative
(positive) local shear.
}\label{Fpbutter}\end{figure}

The toroidal component of the mean field displays dynamo waves travelling
in opposite directions at different $x$-positions, depending on the local
sign of the shear. For $x=-\pi$ the local shear is negative and the dynamo
wave travels in the positive $z$-direction, whilst for $x=0$ the local
shear is positive and the wave travels in the negative $z$-direction
(at least after $t=4000$); see \Fig{Fpbutter}. This is consistent with
what is predicted from mean-field $\alpha\Omega$ dynamo theory (e.g.\
Yoshimura 1975). The dynamo wave at $x=-\pi$ is quite well established
at $t=2000$, but the behaviour at $x=0$ is more complicated and a clear
dynamo wave develops only after $t=4000$. The cycle period at $x=0$ is
also longer than at $x=-\pi$. This somewhat complicated behaviour suggests
that the turbulence properties may not be homogeneous in $x$, which could
be a consequence of the magnetic feedback.

\epsfxsize=8.9cm

\epsfxsize=8.9cm\begin{figure}\epsfbox{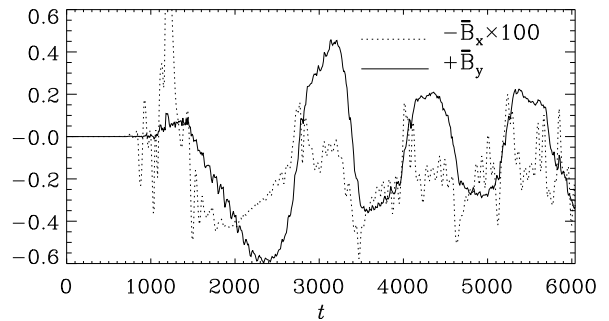}\caption[]{
Evolution of $\overline{B}_x$ and $\overline{B}_y$ at $x=-\pi$ and
$z=0$. Note that $\overline{B}_x$ has been scaled by a factor $-100$.
}\label{Fpbutter_line}\end{figure}

There is a systematic phase shift and a well-defined amplitude ratio
between $B_y$ and $B_x$; see \Fig{Fpbutter_line}. Note also that the
dynamo wave is markedly non-harmonic. These are clear properties that
can be compared with mean-field model calculations (\Sec{mfinterpret}).

\begin{figure}\epsfbox{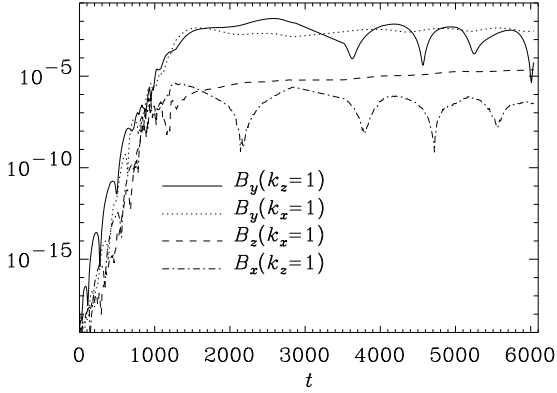}\caption[]{
Evolution of the power, $|\hat{B}_i(k_j)|^2$, of a few selected
Fourier modes. After $t=1700$, most of the power is in the mode
$|\hat{B}_y(k_z)|^2$, i.e.\ in the toroidal field component with
variation in the $z$-direction.
}\label{Fpbmk1}\end{figure}

Before we turn to the saturation of the field at the scale of
the box we first want to assess the relative importance of the different
Fourier modes at different times. Thus, we plot in \Fig{Fpbmk1} the evolution of
the power, $|\hat{B}_i(k_j)|^2$, in a few selected modes. Note that after
$t=1700$, most of the power is in the mode $|\hat{B}_y(k_z)|^2$, i.e.\
the toroidal field component with variation in the $z$-direction. Between
$t=1700$ until $t\approx3500$ the ratio of toroidal to poloidal field
energies is around $10^4$, so $B_{\rm tor}/B_{\rm pol}\approx50$. At
later times this ratio diminishes somewhat. This may suggest that there
is a growing contribution from $\alpha^2$-type dynamo action. This is
also supported by the apparently independent evolution of the oscillatory
$k_z$-mode and the non-oscillatory $k_x$-mode; see \Fig{Fpbmk1}.

\epsfxsize=8.2cm

\epsfxsize=8.2cm\begin{figure}\epsfbox{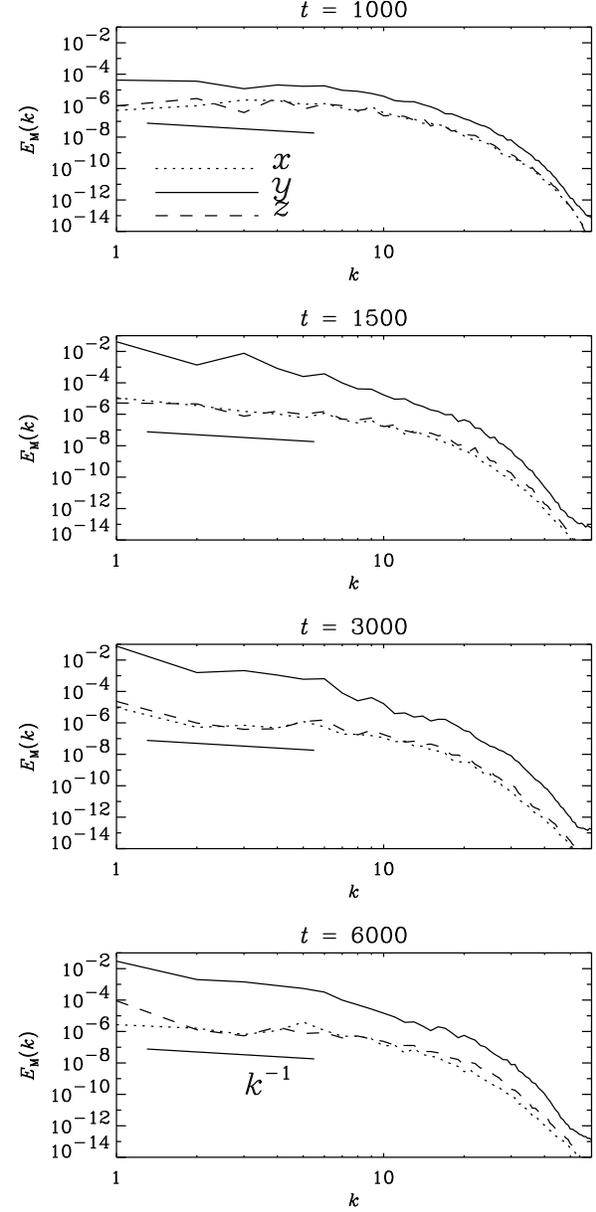}\caption[]{
Two-dimensional power spectra of the three components of the mean field,
$\meanBB_y$ (solid for the $y$ component, and broken lines for the $x$
and $z$ components. The $k^{-1}$ slope is given for comparison.
}\label{Fpspe}\end{figure}

\begin{figure}\epsfbox{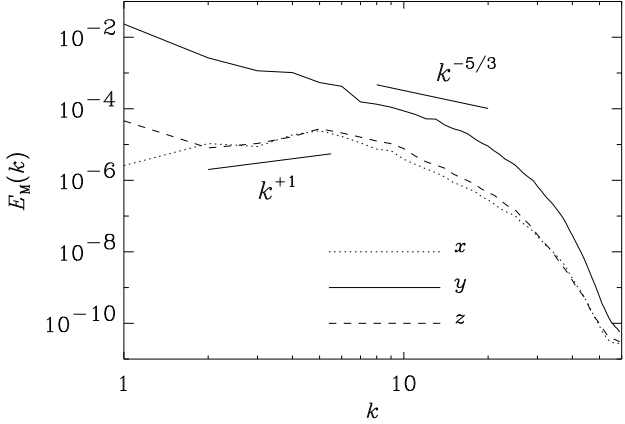}\caption[]{
Three-dimensional power spectrum of the three field components.
$120^3$ meshpoints, $t=6000$.
}\label{Fpspe3d}\end{figure}

In \Fig{Fpspe} we show two-dimensional power spectra of the three
components of the mean field, $\meanBB$. (Here and elsewhere we denote
$y$-averaged fields by a bar whilst angular brackets are used for
full volume averages.) Note that a strong toroidal field builds up first,
and at later times the poloidal field components also gain significant
power at the largest scale (i.e.\ at $\kk^2<2$). One should bear in
mind, however, that these spectra are for the {\it mean} fields. The
three-dimensional power spectra of the non-averaged fields reveal that
the poloidal fields are `noisy' and possess significant power
at the forcing wavenumber, $k_{\rm f}$; see \Fig{Fpspe3d}.

\epsfxsize=8.9cm

\begin{figure}\epsfbox{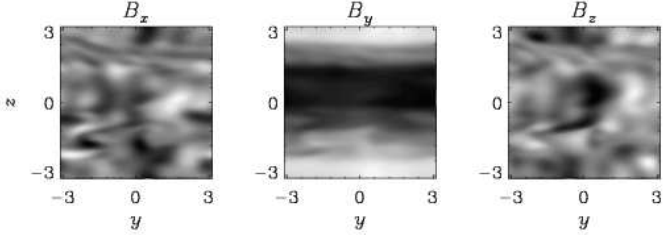}\caption[]{
Images of the three components of $\BB$ in an arbitrarily chosen $yz$ plane.
Note that $B_x$ and $B_z$ show strong variations in $y$, but $B_y$ does not.
$t=6000$.
}\label{Fpbtor}\end{figure}

The small scale contributions to the poloidal field result from variations
in the toroidal direction, as can be seen in a longitudinal cross-section;
see \Fig{Fpbtor}, where we show images of the three field components
in the $yz$ plane. The figure shows that whilst the toroidal field
is relatively coherent in the toroidal direction, the poloidal field
components are much less coherent and show significant fluctuations in
the $y$-direction.

\epsfxsize=8.9cm

We now turn to the temporal evolution of the resulting large scale
magnetic field that gradually emerges during this simulation. We begin
by briefly reviewing the main results in the absence of shear (B2001).

\section{Resistively limited growth on large scales}
\label{Sresist}

In an unbounded or periodic system the magnetic helicity,
$\bra{\AAA\cdot\BB}$, can only change if there is microscopic magnetic
diffusion, $\eta$, and finite current helicity, $\bra{\JJ\cdot\BB}$,
\EQ
{\dd\over\dd t}\bra{\AAA\cdot\BB}=-2\eta\bra{\JJ\cdot\BB}.
\label{helconstr}
\EN
In B2001 a possible configuration for the large scale magnetic field was
\EQ
\meanBB=B_0\pmatrix{\cos(k_1z+\varphi_x)\cr\sin(k_1z+\varphi_y)\cr 0},
\EN
which corresponds to a force-free magnetic field that varies in the
$z$-direction, although variations in one of the other two coordinate
directions, and with arbitrary phase shifts $\varphi_x$ ($\approx\varphi_y$),
were also possible (B2001). $B_0=\bra{\meanBB^2}^{1/2}$ is the amplitude,
whose time dependence was found to be subject to the helicity constraint
(B2001).

The present case is different because of shear which tends to increase the
toroidal field, but not the poloidal field. We model this by writing
\EQ
\meanBB=\pmatrix{B_{\rm pol}\,\cos(k_1 z+\varphi_x)\cr B_{\rm tor}\,
\sin(k_1 z+\varphi_y)\cr 0},
\label{BtorBpol}
\EN
where $B_{\rm pol}$ and $B_{\rm tor}$ are the amplitudes of the poloidal
and toroidal field components. In addition to the $z$-dependence there can
also be an $x$-dependence of the mean field, which is natural due to the
$x$-dependence of the imposed shear profile. However, for the following
argument all we need is the fact that the magnetic and current helicities
are proportional to the product of poloidal and toroidal field magnitudes,
\EQ
\bra{\meanJJ\cdot\meanBB}/k_1\approx\mp B_{\rm tor}\,B_{\rm pol}\approx
k_1\bra{\meanAA\cdot\meanBB}.
\label{meanJB}
\EN
The upper sign applies to the present case where the kinetic
helicity is positive (representative of the southern hemisphere), and the
approximation becomes exact if the field is indeed represented by \Eq{BtorBpol}.

Following B2001, in the steady case $\bra{\AAA\cdot\BB}=\mbox{const}$,
see \Eq{helconstr}, and so the r.h.s.\ of \Eq{helconstr} must vanish,
i.e.\ $\bra{\JJ\cdot\BB}=0$, which can only be consistent with \Eq{meanJB}
if there is a small scale component, $\bra{\jj\cdot\bb}$, whose sign is
opposite to that of $\bra{\meanJJ\cdot\meanBB}$. Hence we write
\EQ
\bra{\JJ\cdot\BB}=\bra{\meanJJ\cdot\meanBB}+\bra{\jj\cdot\bb}\approx0.
\EN
This yields, analogously to B2001,
\EQ
-{\dd\over\dd t}\left(B_{\rm tor}\,B_{\rm pol}\right)
=+2\eta k_1^2\left(B_{\rm tor}\,B_{\rm pol}\right)
-2\eta k_1|\bra{\jj\cdot\bb}|,
\label{approxB_ode}
\EN
which yields the solution
\EQ
B_{\rm tor}\,B_{\rm pol}
=\epsilon_0 B_{\rm eq}^2\left[1-e^{-2\eta k_1^2(t-t_{\rm s})}\right],
\label{approxB}
\EN
where $\epsilon_0=|\bra{\jj\cdot\bb}|/(k_1 B_{\rm eq}^2)$ is a
prefactor, $B_{\rm eq}$ is the equipartition field strength with
$B_{\rm eq}^2=\mu_0\bra{\rho\uu^2}$, and $t_{\rm s}$ is the time when
the small scale field has saturated which is when \Eq{approxB_ode}
becomes applicable. All this is equivalent to B2001, except that
$\bra{\meanBB^2}$ is now replaced by the product $B_{\rm tor}\,B_{\rm
pol}$. The significance of this expression is that large toroidal fields
are now possible if the poloidal field is weak.

\epsfxsize=8.2cm\begin{figure}\epsfbox{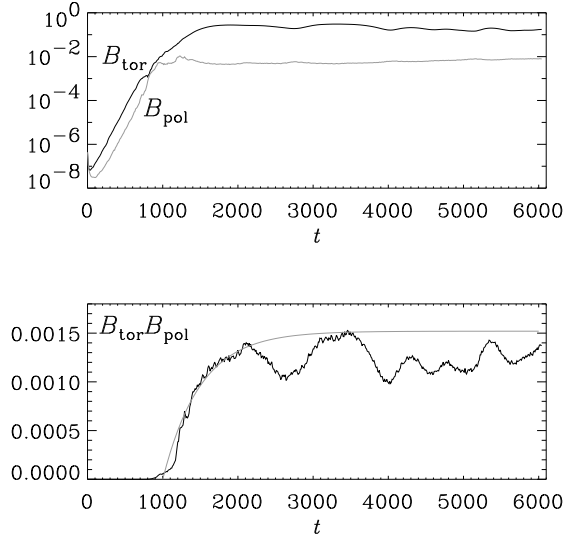}\caption[]{
Growth of poloidal and toroidal magnetic fields on a logarithmic scale
(upper panel), and product of poloidal and toroidal magnetic fields on
a linear scale. For the fit we have used $k_1^2=2$, $B_{\rm eq}=0.035$,
and $\epsilon_0=1.3$.
}\label{Fpbm1}\end{figure}

In order to compare with the simulation we now define
\EQ
B_{\rm tor}\equiv\bra{\meanBB_y^2}^{1/2},\quad
B_{\rm pol}\equiv\bra{\meanBB_x^2+\meanBB_z^2}^{1/2}.
\EN
Note that this definition generalizes that given in \Eq{BtorBpol}. In
\Fig{Fpbm1} we show the evolution of $B_{\rm tor}$ and $B_{\rm pol}$
and compare the evolution of the product $B_{\rm tor}\,B_{\rm pol}$
with \Eq{approxB}. There are different stages; for $1200<t<2200$ and
$3000<t<3700$ the effective value of $k_1^2$ is $k_1^2=2$ (because there
are contributions from $k_x=1$ and $k_z=1$; see \Fig{Fpbmk1}), whilst at
other times ($2500<t<2800$ and $t>4000$) the contribution from $k_x=1$
(for $2500<t<2800$) or $k_z=1$ (for $t>4000$) has become subdominant
and we have effectively $k_1^2=1$. This is consistent with the change
of field structure discussed in the previous section: for $2000<t<3000$
and around $t=4000$ the $B_y(k_x=1)$ mode is less powerful than the
$B_y(k_z=1)$ mode.

We may conclude that the effect of the helicity constraint is clearly
identified in the present simulations. This is substantiated by the
fit shown in \Fig{Fpbm1}. The episodes during which the field amplitude
is below that obtained from the helicity constraint can be explained by
temporary changes in the field geometry.

\section{Astrophysical implications}

The main result of this paper is a quantitative modification of the
helicity constraint for dynamos in the presence of shear. With shear
included the estimate for $\bra{\meanBB^2}$ of B2001 is now to be replaced
by the product $B_{\rm tor}\,B_{\rm pol}\approx\bra{\meanBB^2}/Q$, where
$Q=B_{\rm tor}/B_{\rm pol}\gg1$ and so $\bra{\meanBB^2}\approx B_{\rm
tor}^2$. For {\it early times}, the exponential function in \Eq{approxB}
can be expanded:
\EQ
\bra{\meanBB^2}\approx\epsilon_0 B_{\rm eq}^2 2\eta k_1^2(t-t_{\rm s})Q.
\EN
In the case of efficient large scale dynamo action the small scale
current helicity is very nearly equal to the normalized kinetic helicity,
$\rho_0\bra{\oo\cdot\uu}$ (see also Brandenburg \& Subramanian 2000),
which in turn is approximately $k_{\rm f}\bra{\rho\uu^2}$. Since
$\epsilon_0=\bra{\jj\cdot\bb}/(k_1 B_{\rm eq}^2)$, this leads to
$\epsilon_0\approx k_{\rm f}/k_1$, which is 5 in the present case.
The value of $\epsilon_0$ that fits the simulation results best is only
1.3 (see \Fig{Fpbm1}), so the dynamo seems to be not fully efficient.
This reduced efficiency could partly be explained by the fact that the
actual field is not sinusoidal, as assumed in \Eq{BtorBpol}, and that
the phase shift between poloidal and toroidal fields in not optimal.

We now want to estimate the time, $\tau_{\rm eq}$, required to build up
a large scale field of equipartition field strength, i.e.\
$\bra{\meanBB^2}=B_{\rm eq}^2$. In units of the turnover time,
$\tau=L/u_{\rm rms}$, we have
\EQ
\tau_{\rm eq}/\tau
=u_{\rm rms}L/(2\eta k_1^2L^2\epsilon_0 Q)
=R_{\rm m}/(\epsilon_1 Q),
\label{ratio}
\EN
where we have introduced a new coefficient $\epsilon_1=2\epsilon_0
k_1^2L^2$. Applying this to the sun we have $\tau_{\rm
eq}/\tau\approx10^4$--$10^7$, if we assume $R_{\rm m}=10^8$--$10^{10}$,
$Q=10-100$, and $\epsilon_1\approx2(2\pi)^2\approx100$.
At the bottom of the solar convection zone the turnover time is about
10 days ($0.03\yr$), so the time scale for building up a large scale
field to equipartition strength is between 300 and $3\times10^5\yr$.

We have not yet calculated models with different values of the magnetic
Reynolds number, so we cannot properly assess the effect on the cycle
period. If the cycle period scales in the same way as the growth time
of the dynamo, then the helicity constraint would, even in the presence
of shear, continue to pose a serious problem for understanding cyclic
activity of solar-like stars.
However, before making more detailed comparisons with astrophysical bodies
it would be important to assess the importance of open boundaries,
for example. This seems to be now one of the most important remaining
aspects to be clarified in the theory of large scale dynamos; see also
Blackman \& Field (2000) and Kleeorin \ea (2000). Although initial
results from simulations with open boundaries seem pessimistic in
that respect (Brandenburg \& Dobler 2001), the effects of open
boundaries are likely to be more important in cases with outflows
(e.g.\ in protostellar accretion discs or in active galactic nuclei).
It should also be mentioned that large scale dynamos may operate with
non-helical flows; see the recent papers by Vishniac \& Cho (2000) and
Zheligovsky \ea 2000). This
may relax the helicity constraint, but so far there are no simulations
supporting this possibility.

\section{Mean-field interpretation}
\label{mfinterpret}

In the absence of shear the results of the simulations could be modelled
quite well in terms of a mean-field $\alpha^2$-dynamo with simultaneous
quenching of the $\alpha$-effect and the turbulent diffusivity. In this
section we shall try to do the same for the $\alpha\Omega$-dynamo. Since the
shear is strong compared with the inverse turnover time we can make the
$\alpha\Omega$-approximation, i.e.\ we can neglect the $\alpha$-effect
in the equation for the generation of the toroidal magnetic field. We
also assume that the magnetic field varies only in the direction of
the vorticity vector of the shear, which is the direction in which the
dynamo wave travels (Yoshimura 1975). In the present case this is the
$z$-direction. Thus, the relevant equations, in terms of the mean
vector potential $\meanAA$, are
\EQ
\partial_t\overline{A}_x=-S\overline{A}_y
+\eta_{\rm T}\partial_z^2\overline{A}_x,
\label{dAx}
\EN
\EQ
\partial_t\overline{A}_y=+\alpha\overline{B}_y
+\eta_{\rm T}\partial_z^2\overline{A}_y,
\label{dAy}
\EN
where $\overline{B}_y=\partial_z\overline{A}_x$ and
$\eta_{\rm T}=\eta+\eta_{\rm t}$ is the {\it total} (microscopic plus
turbulent) magnetic diffusivity. [In \Eq{dAx} we have used a particular
gauge that allowed us to write the shear term as $S\overline{A}_y$;
see Brandenburg \ea (1995) for details.]  As in the case of the
$\alpha^2$-dynamo, we shall assume that $\eta_{\rm t}$ and $\alpha$
are quenched in the same way:
\EQ
\alpha={\alpha_0\over1+\alpha_B\meanBB^2\!/B_{\rm eq}^2},\quad
\eta_{\rm t}={\eta_{\rm t0}\over1+\eta_B\meanBB^2\!/B_{\rm eq}^2},
\label{quench}
\EN
where $\alpha_B=\eta_B$ is assumed, and
$\meanBB^2=\overline{B}_x^2+\overline{B}_y^2$ with
$\overline{B}_x=-\partial_z\overline{A}_y$.
In the case of the $\alpha^2$-dynamo
in a periodic domain the two components of the magnetic field were
sinusoidal and phase shifted by $90^\circ$ such that $\meanBB^2$ was
spatially constant. It was therefore possible to obtain the solution
for the evolution of $\meanBB^2$ in closed form. The final saturation
field strength, $B_{\rm fin}$, was then given by [Eq.~(55) of B2001]
\EQ
\alpha_B{B_{\rm fin}^2\over B_{\rm eq}^2}\approx{\lambda\over\eta k_1^2}
\quad\mbox{(for the $\alpha^2$-dynamo)},
\label{alpBper}
\EN
where $\lambda=\alpha_0 k_1-\eta_{\rm T0}k_1^2$ is the kinematic growth
rate of the $\alpha^2$-dynamo.

\begin{figure}\epsfbox{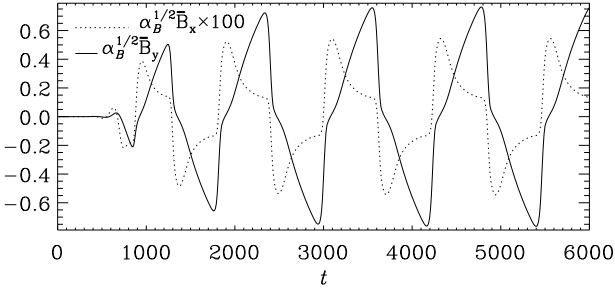}\caption[]{
Evolution of $\alpha_B^{1/2}\overline{B}_x$ and
$\alpha_B^{1/2}\overline{B}_y$ in the
one-dimensional mean-field model with ${\cal D}=10$, $\lambda=0.015$
and $\eta=5\times10^{-4}$. Note that $\overline{B}_x$ has been scaled
by a factor $100$. (In this case $S>0$, so we have plotted $+B_x$,
and not $-B_x$ as we did in \Fig{Fpbutter_line} where $S<0$.)
}\label{Fptime}\end{figure}

In the present case of an $\alpha\Omega$-dynamo, $\meanBB^2$ is no
longer spatially constant and the solution cannot be obtained in closed
form. We therefore resort to numerical solutions of \Eqss{dAx}{quench}
using periodic boundary conditions. All the solutions turned out to be
oscillatory with a period $T$, but the temporal structure is strongly
non-harmonic; see \Fig{Fptime}. Note that the time dependence of
$\overline{B}_x$ and $\overline{B}_y$ is qualitatively and quantitatively
similar to that found in the actual simulation (\Fig{Fpbutter_line}).
The field amplitude depends on the value of $\alpha_B$ and agrees with
that found in the simulation (\Fig{Fpbutter_line}) if $\alpha_B\approx2$.
The solution also depends on the value of the dynamo number
\EQ
{\cal D}=\alpha_0 k_1 S/(\eta_{\rm T0}k_1^2)^2,
\EN
where $\eta_{\rm T0}=\eta+\eta_{\rm t0}$ is the kinematic value
of the total turbulent magnetic diffusivity. For the model shown in
\Fig{Fptime} we used ${\cal D}=10$, but if ${\cal D}$ is doubled
the cycle period also approximately doubles. Thus, ${\cal D}=20$
would be more representative for the dynamo wave at $x=0$
(cf.\ \Fig{Fpbutter}).

\epsfxsize=8.2cm

Although the present analysis is straightforward and
indeed quite similar to other $\alpha\Omega$-dynamos considered
in the literature (e.g.\ Moffatt 1978, Krause \& R\"adler 1980),
a main conceptual difference is that we consider here $\alpha$
and $\eta_{\rm t}$ to be quenched in the same way, and that we retain
the microscopic magnetic diffusion $\eta$ which is not quenched.

\begin{figure}\epsfbox{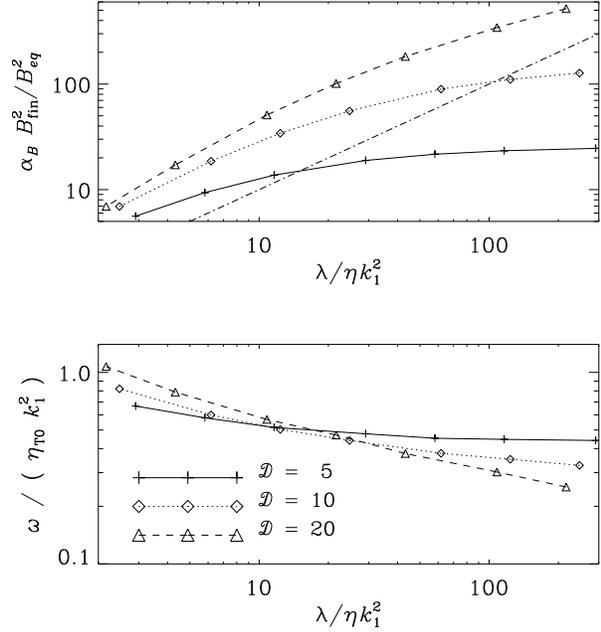}\caption[]{
Normalised saturation field strength and cycle frequency for the
saturated state of a nonlinear one-dimensional $\alpha\Omega$-dynamo
with simultaneous $\alpha$- and $\eta_{\rm t}$-quenchings.
The diagonal (dash-dotted line) in the first panel gives the result for
the corresponding $\alpha^2$-dynamo (for all values of
$\alpha_0/\eta_{\rm T0}k_1$).
}\label{Fpres}\end{figure}

We have determined the value of $\alpha_B B_{\rm fin}^2/B_{\rm eq}^2$
and the cycle frequency $\omega=2\pi/T$ as a function of
$\lambda/\eta k_1^2$ for different values of the dynamo number
${\cal D}$. The results are shown in
\Fig{Fpres}. We have checked that the different curves in \Fig{Fpres}
depend only on the parameter ${\cal D}$, regardless of the values of
$\alpha_0$, $S$ and $\eta_{\rm T0}$, provided the kinematic growth rate
of the linearized form of \Eqs{dAx}{dAy},
\EQ
\lambda=-\eta_{\rm T0}k_1^2+\sqrt{\alpha_0 k_1 S/2},
\label{lambda_ao}
\EN
is kept unchanged. Note that it is now this $\lambda$ that enters in
the expression
$\lambda/\eta k_1^2$, which we have considered as the control parameter
for the numerical solutions displayed in \Fig{Fpres}.

With these preparations we can now make a detailed comparison with
the simulation data. In the simulation the kinematic growth rate
can be read off the first panel of \Fig{Fpbm1} and turns out to be
$\lambda=0.015$. Thus, with $\eta=5\times10^{-4}$ and $k_1=1$ we have
$\lambda/\eta k_1^2=30$. From \Fig{Fpres} we see, then, that
$\omega/\eta_{\rm T0}k_1^2\approx0.4$. For $T$ in the range
1000--2000 we have $\omega=0.006$--$0.003$, which yields
$\eta_{\rm T0}\approx0.015$--$0.0075$, respectively.

\epsfxsize=8.2cm

Given the values of $\lambda$ and $\eta_{\rm T0}k_1^2$, we can express
the dynamo number as
\EQ
{\cal D}/{\cal D}_{\rm crit}=\left[\lambda/(\eta_{\rm T0}k_1^2)+1\right]^2,
\EN
where ${\cal D}_{\rm crit}=2$ is the critical value for dynamo action,
and find ${\cal D}=8$--$18$ for $T=1000$--$2000$, respectively. From \Fig{Fpres}
we see, then, that $\alpha_{\rm B}B_{\rm fin}^2/B_{\rm eq}^2=60$--$100$.
In the simulations we have $B_{\rm fin}\approx0.25$ and
$B_{\rm eq}=0.035$, so $B_{\rm fin}^2/B_{\rm eq}^2\approx50$, and
therefore $\alpha_{\rm B}=1$--$2$, which is in agreement with the value
inferred earlier from the field amplitude; cf.\ \Figs{Fpbutter_line}{Fptime}.
We can therefore conclude that in an $\alpha\Omega$-dynamo, $\alpha$
and $\eta_{\rm t}$ are quenched much less than in an $\alpha^2$-dynamo,
where $\alpha_{\rm B}$ would be around 30.
If the weaker quenching for oscillatory $\alpha\Omega$-type dynamos is
confirmed for larger values of the magnetic Reynolds number this would
also suggest that the cycle period is also only weakly increased.
Already now the cycle period is closer to the dynamical time scale
than to the resistive one.
This is best seen by comparing the two ratios
\EQ
\omega/\eta_{\rm T0}k_1^2\approx0.4\quad\mbox{and}\quad
\omega/\eta k_1^2\approx6\mbox{--}12.
\EN
Note also that the values of $\eta_{\rm T0}$ and $\lambda$ are very
close to each other. This confirms again that the turbulent diffusivity
is dynamically significant and not quenched to its microscopic value.

\epsfxsize=8.2cm\begin{figure}\epsfbox{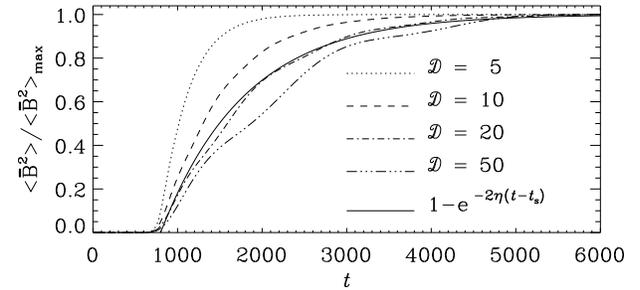}\caption[]{
Resistively dominated saturation behaviour in the $\alpha\Omega$-dynamo
for large enough dynamo numbers (${\cal D}\approx20$). For all curves
we have $\lambda=0.015$ and $\eta=5\times10^{-4}$. For large values of
${\cal D}$ the cycle oscillation begin to distort the curve and
cause additional deviations from the helicity constraint (solid line),
which is best matched for ${\cal D}=20$.
}\label{Fpsat}\end{figure}

Finally we show in \Fig{Fpsat} the evolution of
$\bra{\meanBB^2}/\bra{\meanBB^2}_{\max}$ for different values of
${\cal D}$, and compare with the shape of the curve predicted by the
helicity constraint of \Eq{approxB}. We see that the correct shape of the
helicity constraint is matched for ${\cal D}\approx20$, which corresponds
to the value obtained for $T=2000$. The fact that the helicity constrained
is matched for one particular value of ${\cal D}$ is surprising.
This suggests that in nonlinear $\alpha\Omega$ dynamo theory the dynamo
number is no longer a free parameter, and that there is only one possible
value of ${\cal D}$ for which the helicity constraint with the correct
value of the {\it microscopic} magnetic diffusivity can be matched.

\section{Conclusions}

The present investigations have shown that the effects of the helicity
constraint can clearly be identified, even though much of the field
amplification results now from the shearing of a poloidal field. Instead
of having a constraint on the magnetic energy in the mean field,
one now has a constraint
on the geometrical mean of the energies in the poloidal and toroidal mean
field components. The dynamo remains time dependent with a typical period
that is closer to a dynamical time scale than to a resistive one.
The toroidally averaged field alternates in sign
and shows a clear migration pattern.

The present work has revealed that, even though the kinetic helicity of
the flow is near its maximum possible value, the poloidal field shows
a great deal of `noise', whilst the toroidal field does not. Power spectra
of the poloidal field show that most of the power is in small scales,
making the use of averages at first glance questionable. However, once
the field is averaged over the toroidal direction the resulting poloidal
field is governed by large scale patterns (the slope of the spectrum is
steeper than $k^{-1}$, which is the critical slope for equipartition
of energy between small and large scale fields). The presence even of
a weak mean poloidal field is crucial for understanding the resulting large
scale field generation in the framework of an $\alpha\Omega$-dynamo.

The results of the simulations can be reproduced by a mean-field
$\alpha\Omega$-dynamo where alpha-effect and turbulent magnetic
diffusivity are quenched by the magnetic field. The strength of the
quenching is however much weaker than for the corresponding
$\alpha^2$-dynamo. The resistively limited growth imposed
by the magnetic helicity constraint is recovered for one particular
value of the dynamo number. Whether or not the cycle period becomes
catastrophically long in the limit of large magnetic Reynolds numbers is
not entirely clear, because the frequency dependence shown in \Fig{Fpres}
seems to level off at a definite value. However, the value of the magnetic
Reynolds number where the cycle frequency levels off shifts to larger
values as the dynamo number is increased. If it is confirmed that large
magnetic Reynolds numbers (based on the microscopic magnetic diffusivity)
also imply large dynamo numbers (based on the value of the turbulent
magnetic diffusivity), then the cycle period would probably be too long
to explain the cycle periods observed in many late type stars.
On the other hand, using a mean-field model that obeys the magnetic
helicity constraint we found evidence that the cycle period is
controlled primarily by the dynamical time scale.

It is important to remember that the flows considered in the
present investigations are driven by some imposed body force.
In astrophysical bodies the flows are driven by convection and shear.
This does not directly affect the helicity constraint which controls the
long time scales discussed here. However, when open boundary conditions
are considered it may be possible that real astrophysical flows have a
better ability to dispose small scale fields whose magnetic helicity has
the opposite sign of that of the large scale field. [In externally driven
flows, open boundaries do not seem to relax sufficiently the constraint
imposed by helicity conservation; see Brandenburg \& Dobler (2001).]

The driven flows considered here and in related papers have the tremendous
advantage of allowing significant progress to be made in understanding
the simulation results quantitatively in terms of mean-field theory.
This will be a much harder task for real astrophysical flows.
For example, the helicity constraint has to our knowledge never been
identified in simulations of astrophysically driven flows.
This seems to be now one of the most important tasks for future
simulations of large scale dynamos.

\section*{Acknowledgments}
We thank Eric Blackman and Anvar Shukurov for many stimulating discussions
and comments on
the manuscript. We also thank an anonymous referee for making useful
suggestions that have led to an improved presentation of the results.
ABi and KS acknowledge Nordita for hospitality during
the course of this work. Use of the PPARC supported supercomputers in
St Andrews and Leicester is acknowledged.

\end{document}